# Assessing Excess Mortality in Times of Pandemics Based on Principal Component Analysis of Weekly Mortality Data– The Case of COVID-19

**Patrizio Vanella\*,[1,2], Ugofilippo Basellini[3], Berit Lange[1,2,4]**

\* Corresponding author: patrizio.vanella@helmholtz-hzi.de

## Declarations

### Availability of data and materials
The datasets used and/or analyzed during the current study are available from the corresponding author on reasonable request.

### Competing interests
The authors declare that they have no competing interests.

### Funding
PV and BL received funding from the European Union's Horizon 2020 research and innovation program under grant agreement No 101003480 and from the Initiative and Networking Fund of the Helmholtz Association.

### Authors' contributions
PV conceptualized the study, structured the data, developed the model, ran the simulations, and wrote the raw version of the text. UB organized the underlying data and discussed the analysis results. PV and UB illustrated the study results. BL contributed to the development of the idea and design of the paper. UB and BL discussed the modeling approach. All authors participated in the literature research. All authors revised the text and agreed to its submission.

### Acknowledgments
We appreciate the helpful comments on an earlier version of the paper by Alexander Kuhlmann from the Center for Health Economics Research Hannover.

---

[1] Department of Epidemiology, Helmholtz Centre for Infection Research (HZI), Inhoffenstr. 7, DE-38124 Brunswick

[2] Hannover Biomedical Research School (HBRS), Carl-Neuberg-Str. 1, DE-30625 Hannover

[3] Laboratory of Digital and Computational Demography, Max Planck Institute for Demographic Research (MPIDR), Konrad-Zuse-Str. 1, DE-18057 Rostock

[4] German Center for Infection Research (DZIF), Inhoffenstr. 7, DE-38124 Brunswick

# Abstract

The current outbreak of COVID-19 has called renewed attention to the need for sound statistical analysis for monitoring mortality patterns and trends over time. Excess mortality has been suggested as the most appropriate indicator to measure the overall burden of the pandemic on mortality. As such, excess mortality has received considerable interest during the first months of the COVID-19 pandemic.

Previous approaches to estimate excess mortality are somewhat limited, as they do not include sufficiently long-term trends, correlations among different demographic and geographic groups, and the autocorrelations in the mortality time series. This might lead to biased estimates of excess mortality, as random mortality fluctuations may be misinterpreted as excess mortality.

We present a blend of classical epidemiological approaches to estimating excess mortality during extraordinary events with an established demographic approach in mortality forecasting, namely a Lee-Carter type model, which covers the named limitations and draws a more realistic picture of the excess mortality. We illustrate our approach using weekly age- and sex-specific mortality data for 19 countries and the current COVID-19 pandemic as a case study. Our proposed model provides a general framework that can be applied to future pandemics as well as to monitor excess mortality from specific causes of deaths.

**Keywords:** COVID-19 Pandemic; Excess Mortality Assessment; Mortality Forecasting; International Mortality Trends; Principal Component Analysis; Time Series Analysis; Monte Carlo Simulation; Stochasticity; Demography; Epidemiology

# 1. Introduction

The current outbreak of COVID-19 has highlighted the need for sound and timely statistical analysis and monitoring of mortality patterns and trends. On many occasions, excess mortality – the number of deaths above expectation in the absence of exceptional events, e.g., a pandemic, an exceptional influenza season, or a heatwave – is considered the most appropriate indicator to measure the overall burden of the pandemic on mortality (National Academies of Sciences & Medicine, 2020). As such, excess mortality due to COVID-19 has received considerable attention during the last few months, including major outlets tracking this indicator across countries (see, e.g., The Economist, 2020). The computation of this measure should be based on long time series of weekly or monthly mortality data, and the COVID-19 pandemic has stimulated the demand for timely release and publication of such data by national authorities (Leon et al., 2020). While aggregate all-cause mortality data are being increasingly released, timely reporting of cause-specific data by demographic subgroups is still underdeveloped. However, such information would allow for near real-time assessment of excess mortality caused by a specific disease. Moreover, widespread statistical approaches for estimating excess mortality have been rather simplistic thus far, as they rely on rather basic statistical measures, which do not include stochasticity or the full dimension of the mortality development into account. Strong correlations in mortality trends exist among not only different demographic groups, but also among adjacent countries (Vanella, 2017); the underlying factors driving mortality reductions, such as advances in medical care and hygiene, reach all of these groups to some extent (Vanella & Deschermeier, 2020). Therefore, a holistic approach for excess mortality assessment should analyze the multitude of demographic and geographic groups simultaneously. Furthermore, rather short time series are generally considered in the computation of excess mortality, which cannot sufficiently capture eventual long-term trends.

In this article, we propose a stochastic framework for estimating excess mortality based on a Lee-Carter modeling approach. We develop a comprehensive model that can take into account the multidimensionality (and eventual collinearity) of the data analyzed, which consists of several long-term time series for 19 different countries, both sexes and four age groups. This allows us to consider the long-term mortality trends present in the data. Finally, the model can produce probabilistic statements concerning the excess mortality that occurs during a particular event. Whereas our case study is on the current COVID-19 pandemic, our method provides a general framework for future outbreaks of other pathogens, as well as other major events influencing mortality on a larger demographic and/or geographic scale.

The following section provides a literature review starting with current approaches of excess mortality estimation in general and specifically during the COVID-19 pandemic. Then we give an overview of stochastic mortality models, which address multiple populations in parallel. The latter gives a theoretical basis for our multidimensional mortality forecast model which is presented in Section 3. Based on this, a stochastic investigation of excess mortality during the COVID-19 pandemic by country and demographics is conducted, whose results are presented in Section 4. These results are then discussed, alongside their implications for mortality forecasting. We finally draw conclusions from our findings and give an outlook on further need for developments in excess mortality evaluation and mortality forecasting.

## 2. Literature Review

## 2.1 Assessment of Excess Mortality and Estimates for COVID-19

Estimation of excess mortality goes back to studies on influenza and pneumonia by Collins et al. for the United States (Collins, 1932; Collins, Frost, Gover, & Sydenstricker, 1930). The authors calculated weekly expected death rates due to influenza, pneumonia, and other causes for the whole population[5] as the median of a seven-year baseline period. These were then compared to the observed death rates due to these causes during a certain period of an epidemic. Positive differences between the observed and the expected mortality rates were then defined as excess mortality. Serfling extended this approach by fitting parametric Fourier (i.e. trigonometric) models to death rate time series separately by age groups and by selected causes of death[6] for estimating monthly excess mortality (Serfling, 1963). Housworth and Langmuir proposed a stochastic extension of Serfling's approach, assuming that the residuals between observed and expected death rates follow a t distribution (Housworth & Langmuir, 1974). Foppa and Hossain proposed a Bayesian extension of that model for excess death numbers due to influenza (Foppa & Hossain, 2008).

Some approaches have been proposed to estimate excess mortality due to COVID-19 during the pandemic. We present some already published results in the form of scientific publications and official reports here.

Magnani et al. estimated expected mortality rates and daily death numbers for Italian regions from January 1st to April 15th, 2020 by averaging the daily mortality rates for the years 2015-2019. Assuming that death counts follow a Poisson distribution, they also estimated 95% confidence intervals (CIs) for the daily death numbers, separately for age groups below 60 years and 60 years and above. They derived a statistically significant excess mortality in Italy due to COVID-19 since March 7th until the end of the study period, estimating 45,032 mean excess deaths, a figure that is more than double the death numbers attributed to COVID-19 (Magnani, Azzolina, Gallo, Ferrante, & Gregori, 2020). Michelozzi et al. further showed that this excess mortality was concentrated in the north of Italy, which has been hit harder by the pandemic than the center and the south of the country, and that excess mortality is more accentuated for men and the elderly (Michelozzi et al., 2020).

---

[5] Census estimates.
[6] Pneumonia or influenza; cardiovascular or renal; others.

The New York City Department of Health and Mental Hygiene (DOHMH) COVID-19 Response team used the regression model of the US Centers for Disease Control and Prevention (CDC) based on the years 2015-2019 to estimate expected deaths. Overall deaths above that margin were defined as excess deaths. This surveillance system is normally applied to estimate excess deaths because of influenza (Centers for Disease Control and Prevention, 2019), but it was further used to estimate excess deaths due to COVID-19 in New York City between March 11[th] and May 2[nd], 2020. The authors identified over 24 thousand excess deaths over the observed period, from which close to 14 thousand were laboratory-confirmed COVID deaths, while other 5 thousand were probably associated with COVID-19 (New York City Department of Health and Mental Hygiene Covid-19 Response Team, 2020).

EUROMOMO concurrently offers excess mortality estimates based on a fit Generalized Linear Model (GLM) by age groups for 23 European countries. The model is fitted to a maximum of the previous five years of data. The graphs reported by EUROMOMO show a general pattern of excess mortality in Europe since April 2020 for individuals aged 15 and older (Statens Serum Institut, 2020a, 2020b).

These approaches do not include correlations in mortality rates among age groups and/or among countries. Our proposed approach aims at overcoming these limitations employing a demographic perspective. Appendix C provides a summary comparison of the presented approaches and results including ours.

## 2.2 Multi-Population Stochastic Mortality Forecasting

There is a large amount of literature on mortality forecasting approaches. As it is not our intention to give a full literature review here, the interested readers are referred to, e.g., the compilation by Janssen (Janssen, 2018). We will restrict our review to those approaches we believe are important in this context, which are stochastic models that include age-specific mortality and multiple populations.

One forecast approach of major importance is based on principal components (PCs). A PC is a linear combination of a group of variables, in our context age-specific mortality rates. The PCs are derived by singular value decomposition. The method has two major advantages. First, highly dimensional phenomena, such as mortality rates among age groups, sex, and countries, can be analyzed relatively efficiently. Second, correlations among different variables, such as age- and sex-specific mortality rates are included in the analysis, which is very important in forecasting to quantify the uncertainty of the mortality forecast adequately. An illustrative explanation of the method applied to age- and sex-specific survival rates is given, e.g., by Vanella (Vanella, 2018). The application of principal component analysis (PCA) to age-specific mortality rates goes back to Ledermann and Breas (Ledermann & Breas, 1959), who used it for transforming French data to derive common mortality trends. Le Bras and Tapinos (Le Bras & Tapinos, 1979) proposed using PCA to project mortality in France. Bell and Monsell (Bell & Monsell, 1991) extended this framework by including autocorrelations of the PCs employing autoregressive integrated moving average (ARIMA) models[7] for mortality forecasting in the US. Lee and Carter (Lee & Carter, 1992) identified the first PC in that model as a general mortality index, which covers the vast majority of mortality trends observed over all age groups, and proposed a random walk with drift model to forecast the index, which can then be retransformed to derive forecast mortality rates. Tuljapurkar et al. (Tuljapurkar, Li, & Boe, 2000) qualitatively showed that there were large correlations in mortality trends among the G7 countries, which could be covered well by the Lee-Carter model. Booth et al. (Booth, Maindonald, & Smith, 2002) proposed a graphical method for determining the optimal baseline period to inform the model. A too-short baseline assumes the long-term future trends to follow the near past, which appears unrealistic. On the other hand, very long past data may not apply to future trends, especially in the shorter term. Whereas the mortality index is modeled as a linear process, Brouhns et al. (Brouhns, Denuit, & Vermunt, 2002) proposed a GLM version of the Lee-Carter model. The classical Lee-Carter model assumes independence between mortality of females and males, which can be rejected (see, e.g., Vanella, 2017, on the correlation of mortality among both sexes). Li and Lee (Li & Lee, 2005) therefore proposed an extension, the so-called common factor model, which includes the international correlations and the correlations between the two sexes in the mortality trends to some degree in the analysis. Renshaw and Haberman (Renshaw & Haberman, 2006) enhanced the classical Lee-Carter model by including cohort effects in the model. Hyndman and Ullah (Hyndman & Ullah, 2007) proposed a nonparametric extension of the Lee-Carter model. Russolillo et al. (Russolillo, Giordano, & Haberman, 2011) proposed extending the Lee-Carter model by applying a Three-Mode PCA to include international correlations in the model. However, they ignored sex-specific differences in their model. Vanella (Vanella, 2017) proposed a simulation approach, which similarly forecasts age- and sex-specific survival rates for 18 European countries, therefore taking correlations in mortality trends among age groups, both sexes and countries into account via PCA, extrapolating long-term trends in the PC time series parametrically and considering stochasticity by Monte Carlo simulation of ARIMA models. The author showed an efficient way to include common international trends in mortality in one model, as PCA can cover the majority of the trends, which different countries witness simultaneously. We will use a derivation of that approach for our analysis. Finally, Bergeron-Boucher et al. (Bergeron-Boucher, Canudas-Romo, Oeppen, & Vaupel, 2017) proposed a modification of the Li-Lee model by leveraging age-at-death distributions and compositional data analysis to produce coherent forecasts for 15 Western European countries.

---

[7] See Box, Jenkins, Reinsel, & Ljung, 2016 or Shumway & Stoffer, 2016 for detailed presentations of ARIMA models.

We see that, with a few exceptions, works on mortality forecasting focused on a national level. In some cases, international or even global mortality forecasts are of interest. Separate forecasts would not only be unfeasible, but would also ignore common trends among the countries. Some authors have conducted stochastic projections of groups of countries or on a global scale using Bayesian approaches, which assume an a priori distribution for some parameter or variable either based on auxiliary data or subjective assumptions (Kruschke, 2015; Lynch, 2007). To capture the major problem of the Lee-Carter model of systematically underestimating the uncertainty in the mortality forecast, Pedroza (Pedroza, 2006) proposed a Bayesian extension of the classical Lee-Carter model, which includes the uncertainty of all parameters using Markov Chain Monte Carlo (MCMC) simulation. King and Soneji (King & Soneji, 2011) suggested considering assumptions on trends in smoking behavior and obesity in projections of age-specific mortality rates through a Bayesian Hierarchical model for the US. Raftery et al. (Raftery, Chunn, Gerland, & Ševčíková, 2013) proposed a Bayesian Hierarchical model for joint probabilistic projections of international male life expectancies by cohort using time series data on life expectancy in combination with judgmental projection data by national experts. The identified distributions are then simulated by MCMC to derive empirical projection intervals for the countries under study. The approach was then expanded for females by simulating the gender gap in life expectancy by regression analysis of the international data (Raftery, Lalić, & Gerland, 2014). The Raftery model is the basis of the life expectancy projections of the United Nations. From these projections, they derive age- and sex-specific mortality rates for all countries with three different techniques, depending on the quality of the mortality data available for the countries (United Nations, 2019). Antonio et al. (Antonio, Bardoutsos, & Ouburg, 2015) give a Bayesian version of the Lee-Carter model which allows for joint mortality projections among various countries.

## 3. Data and Methods

### 3.1 Data

We extracted recently published estimates of weekly mortality rates by sex and age groups below 15 years, 15 to 64 years, 65 to 74 years, 75 to 84 years, and aged 85 and above, provided by the Short-term Mortality Fluctuations data series of the Human Mortality Database (HMD) (Human Mortality Database, 2020). The data give 52 weekly estimates of mortality rates for a series of calendar years, starting from different country-specific time points. We select all countries with available data since the start of the year 2000[8]. We take the data for the whole period week 2, 2000[9] to week 52, 2019 on Austria, Belgium, Estonia, Finland, France, Hungary, Israel, Latvia, Lithuania, the Netherlands, Norway, Poland, Portugal, Scotland, Slovakia, Slovenia, Spain, Sweden, and Switzerland, for which data are available and the population and death numbers are sufficiently high to derive representative estimates of weekly mortality. To avoid zero values in the data, we aggregate the age groups below 15 and 15-64 years in a single group. We compute mortality rates in this wider group using aggregated deaths and exposures extracted from the HMD. Finally, we arrange the data in a 1,039x152 matrix[10] of time series of weekly age-, sex-, and country-specific mortality rates (WASCSMRs). In Appendix A, we report all the country-sex and age-specific combinations analyzed in our paper. For the last step of our analysis, we use daily reported data on COVID-19 associated deaths by country, provided by the European Centre for Disease Prevention and Control (ECDC) (European Centre for Disease Prevention and Control, 2020).

---

[8] With the exceptions of Luxembourg and Iceland, whose relatively small population and death numbers do not allow for derivation of representative weekly estimates.
[9] Scotland does not offer data for Week 1, 2000. To include Scotland in the analysis, we start in week 2 for all countries.
[10] 1,039 weeks in the rows, 4 age groups times 2 sexes times 19 countries in the columns.

## 3.2 Methods

We follow Vanella (2017), which was presented in 2.2. We make sure that later simulations of mortality rates cannot exceed the range (0; 1). This is achieved by using the logistic transformation of the WASCSMRs (Vanella, 2017). We first perform PCA on the logit-WASCSMR time series. Figure 1 illustrates the loadings of the first PC (PC1). The loadings are basically correlations between the PCs and the original variables (Vanella, 2018), in this case, the logit-WASCSMRs.

*Figure 1. Loadings of Principal Component 1*

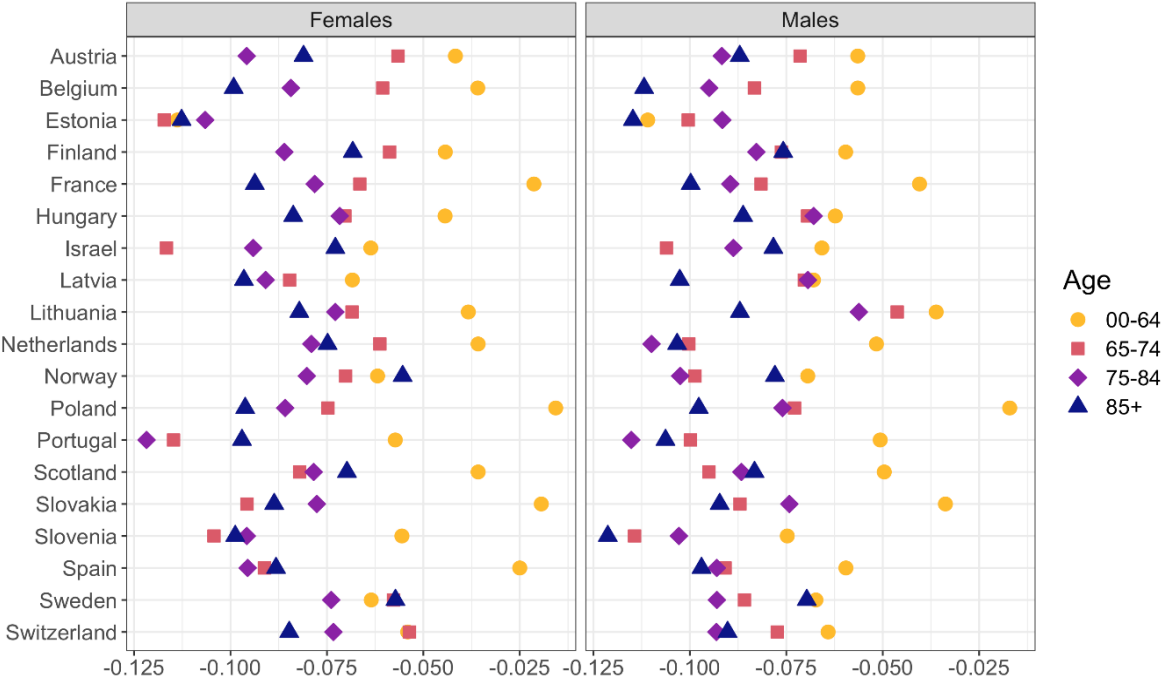

The loadings of PC1 are strictly negative, implying a negative correlation with all mortality rates. Thus, increases in PC1 ceteris paribus are associated with decreases in all WASCSMR under study. PC1 is hence a classical Lee-Carter Mortality Index (Lee & Carter, 1992), explaining 55% of the overall variance in the 152 time series. Therefore, we will address it as *Lee-Carter Index* for the remainder of the paper. Furthermore, it is interesting to observe that the loadings for the younger age group (i.e. 00-64 years) are almost always smaller in absolute value than those for older groups. As such, increases in PC1 generally lead to greater mortality reductions at older rather than younger ages. Figure 2 shows the time series of PC1 from 2000 to 2019. The vertical lines indicate week 1 of each year.

*Figure 2. Past Course of Lee-Carter Index*

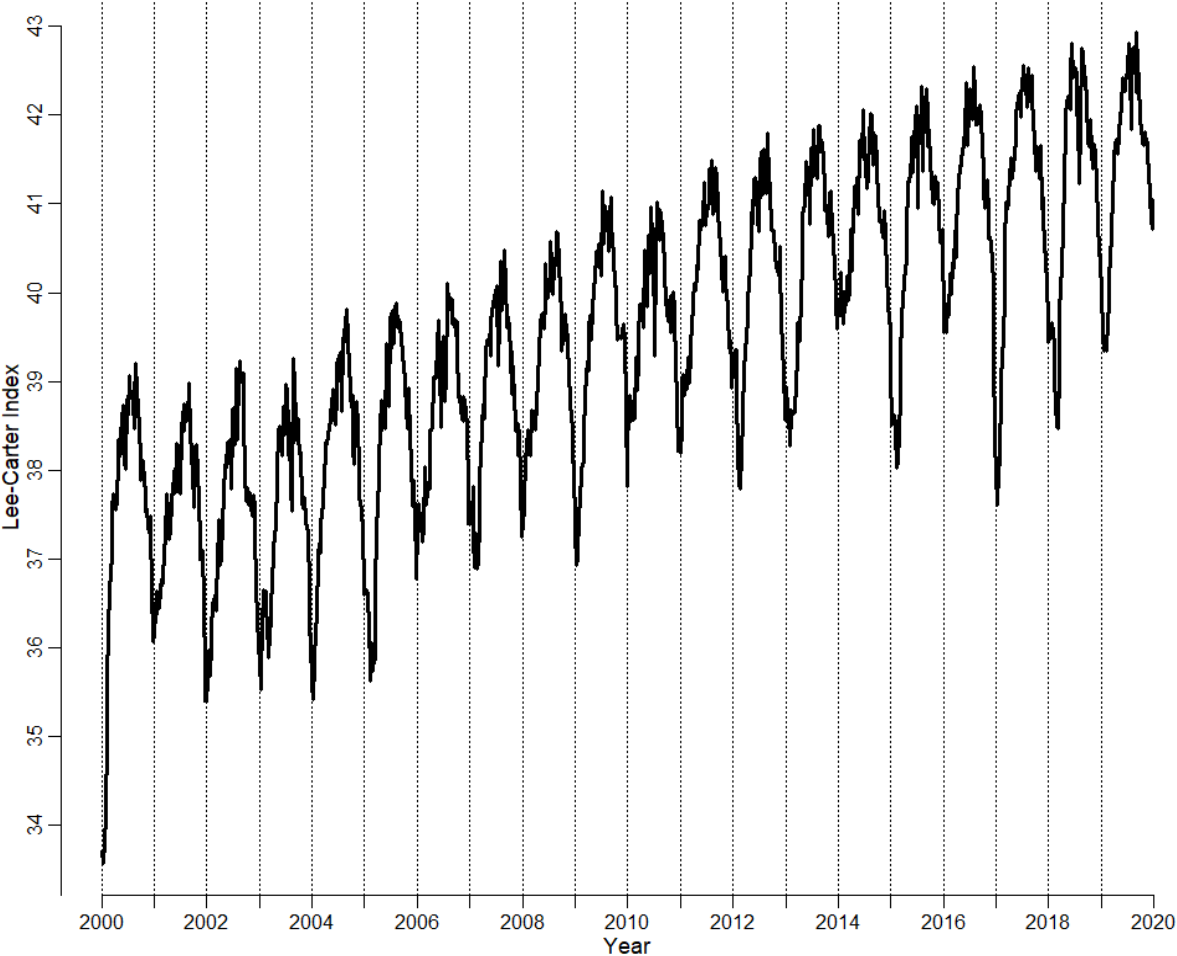

Source: Own computation and design

The curve has a highly seasonal pattern, with strongly increasing mortality (i.e. lower values of PC1) in the winter season and decreasing mortality in summer (i.e. higher values of PC1). The general trend is increasing, which is analog to decreasing mortality trends, but concave, which means that mortality improvements have a diminishing trend. To capture these different features of time series, we iteratively fit a variety of models to the curve, which are then compared via Akaike's Information Criterion (AIC) and the Bayesian Information Criterion (BIC). Table 1 gives the results of this range of model fits.

Table 1. Iterative Trend Function Coefficients with 95% CIs to Lee-Carter Index

| Parameter | 1.1 | 1.2 | 1.3 |
|---|---|---|---|
| **Intercept** | 39.49 (39.4; 39.57) | 33.4 (33.26; 33.54) | 32.95 (32.79; 33.11) |
| $\cos\left(\frac{\pi w}{26}\right)$ | 1.34 (1.22; 1.47) | 1.31 (1.27; 1.35) | 1.05 (0.96; 1.15) |
| $\dfrac{\exp(\frac{w-t_0}{\beta})}{1+\exp(\frac{w-t_0}{\beta})}$ | - | 9.73 (9.51; 9.95) | 9.74 (9.53; 9.94) |
| **Spring** | - | - | 0.71 (0.58; 0.83) |
| **Summer** | - | - | 0.55 (0.36; 0.73) |
| **Autumn** | - | - | 0.5 (0.4; 0.61) |
| *$R^2$* | *0.3037* | *0.9153* | *0.928* |
| *AIC* | *3,707* | *1,520* | *1,358* |
| *BIC* | *3,722* | *1,539* | *1,392* |

The cosine term represents the baseline seasonality of the year, similar to the Serfling-approach, with $w = 0$ being calendar week 31, 2000. This choice of origin leads to the maximization of the $R^2$, indicating the best fit to the observed seasonality. We checked the full Fourier model as well, but discarded the sine term, as it does not lead to any improvement in the model fit, while worsening the efficiency of the model, represented by higher values of the information criteria. The second variable is an inverse logistic growth function, which can be used to simulate a growth function, similar to the implementation in (Vanella, 2017; Vanella & Deschermeier, 2018, 2019) with $w$ being the week and $t_0$ being a parameter to be estimated iteratively to maximize the model's $R^2$. $\beta$ is a parameter, which is computed by Maximum Likelihood Estimation. Spring, Summer, and Autumn are binary variables, which are 1 during the respective seasons and 0 otherwise. Winter is therefore the baseline season. For instance, Spring is from calendar week 13 to 25, Summer from calendar week 26 to 38, and so on. Following Occam's Razor, a simple model should be preferred to a more complex one, if it performs similarly well (Bijak, 2011). A model is most efficient if it minimizes the information criteria. We see that the inclusion of an inverse logistic growth function increases the quality of the model substantially, as it leads not only to a huge increase in the $R^2$ from 30.4 to 91,5%, but also to a large decrease in both the AIC and the BIC. Therefore, the fit of the model to the data increases significantly, while leading to a more efficient model as well. However, this long-term trend is generally not considered in models of excess mortality, as explained in Section 2.1. An extension of the model by seasonal dummies leads to further significant improvements of the fit, as the trigonometric function systematically underestimates the mortality peaks in winter, while overestimating the values in summer. Model 1.3 fits the data well with an $R^2$ of almost 93%. Both information criteria favor this model as well. The coefficients of the seasonal dummies should not be taken for concrete interpretations, however. They simply serve as correction factors to systematic under/over-estimation of the Fourier series. The unexplained share is modeled by fitting a seasonal autoregressive integrated moving average (SARIMA) model, which is chosen by a series of tests following (Vanella, 2018). Figure 3 illustrates the model fit (continuous line) to the data (dots).

*Figure 3. Course of Lee-Carter Index for 2000-2019 with Model Fit*

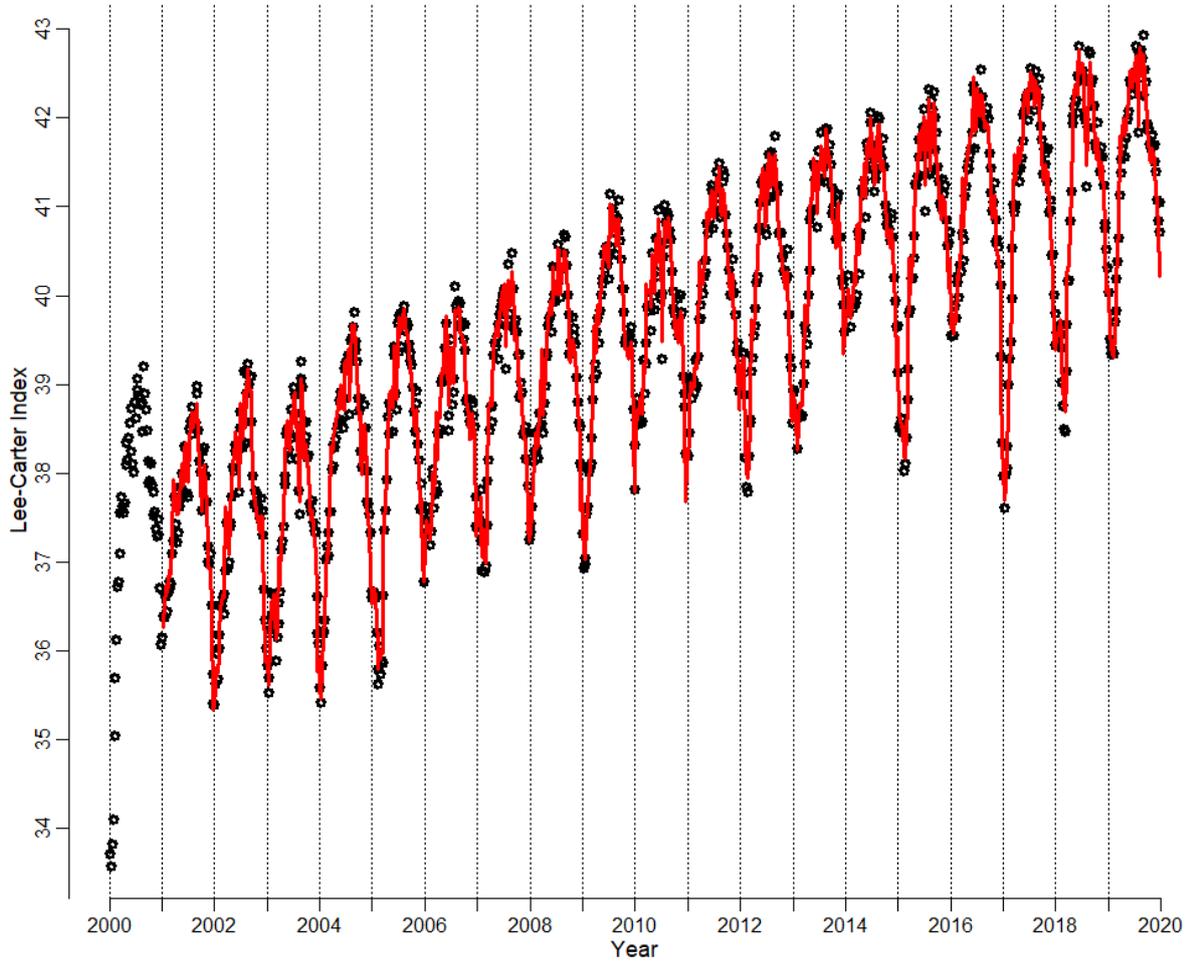

The forecast function underlying the red line mathematically is

$$PC_1(w) = 32.95 + 1.05\cos\left(\frac{w * \pi}{26}\right) + 9.74\frac{\exp\frac{w-220}{482.05}}{1 + \exp\frac{w-220}{482.05}} + 0.71f + 0.55s + 0.5a + \alpha(w),$$

with

- $PC_1(w)$ being the value of the first principal component in week $w$,

- $\alpha(w) = \alpha(w - 1) + 0.16\alpha(w - 52) - 0.16\alpha(w - 53) + \varepsilon(w)$

  $\qquad\qquad -0.26\varepsilon(w - 1)$ , $\varepsilon(w) \sim \mathcal{N}(0; 0.32^2)$,

- $w = 0$ corresponding to calendar week 31, 2000,

- $f$ taking the value 1 in the spring weeks, i.e. calendar weeks 13-25; 0 otherwise,

- $s$ taking the value 1 in the summer weeks, i.e. calendar weeks 26-38; 0 otherwise,

- $a$ taking the value 1 in the autumn weeks, i.e. calendar weeks 39-51; 0 otherwise.

Figure 4 gives the time series with the median forecast from Model 1.3 with theoretical 95% prediction intervals (PIs).

*Figure 4. Historic course of the Lee-Carter Index with Median Forecast and 95% Prediction Intervals*

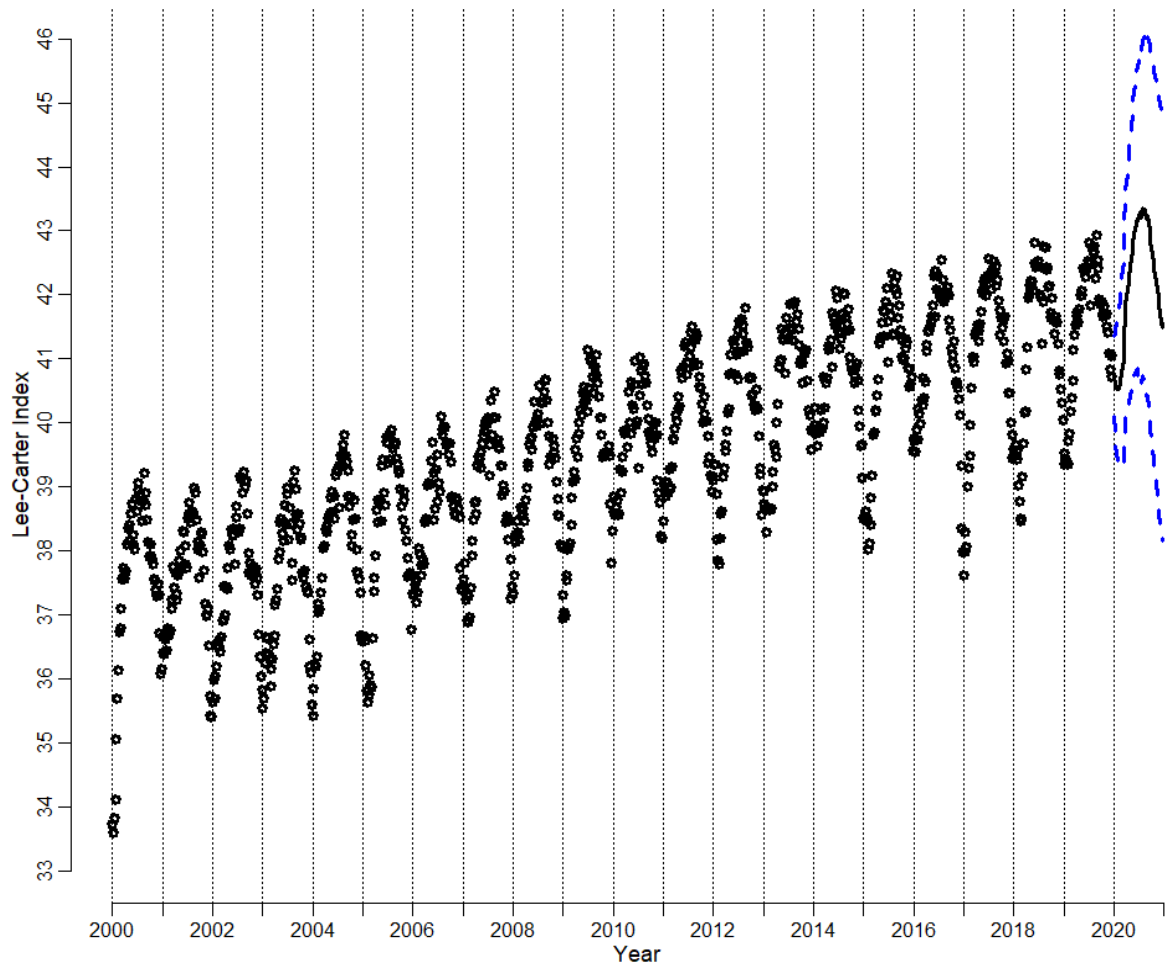

The Lee-Carter Index can serve as a summary indicator of overall mortality, as it captures the main trends in mortality. The loadings of the remaining PCs do not deem as straightforward interpretations and will be assumed random walk processes[11]. Following Vanella (Vanella, 2017), we assume them to follow a random walk process, as our tests show that random walk models perform reasonably well reproducing the series. As the Lee-Carter Index covers general trends in mortality among age groups, sexes, and countries (Lee & Carter, 1992; Vanella, 2017), a comparison of the forecast of its development in 2020 to its observations can give us a general assessment of excess mortality over different groups by week. For this, we multiply the loadings from Figure 1 with the HMD estimates of the WASCSMR in weeks 1-13, 2020 for the study countries, thereby deriving hypothetical observations of the Lee-Carter Index, keeping the loadings derived from the baseline data fixed:

$$\widehat{PC}_1(\tau) = \sum_{i=1}^{152} \lambda_i M_{i,\tau}, \tau = 1,2,\dots,13,$$

With

- $\lambda_i$ being the loading of the i[th] WASCSMR on the Lee-Carter Index,
- $M_{i,\tau}$ being the HMD estimate of the i[th] logit-WASCSMR for week $\tau$.

This will allow a direct comparison between the course of the PC and its expectation based on the time series data. The results of this approach will be presented in Section 4.

We then use Monte Carlo simulation for each PC to simulate 10,000 trajectories of the weekly development of all PCs for the year 2020. Since the PCs are uncorrelated (Vanella, 2018), independent simulation of their future paths does not lead to biased estimation of the mortality rates, which are then derived from these. The results are 10,000 trajectories of each PC, which can be retransformed into weekly trajectories of the logit-WASCSMR. For instance, let $\mathbf{\Pi}_t$ be the simulation matrix of all PCs (10,000x152) in period $t$. The corresponding simulation matrix of the logit-WASCSMRs is then

$$logit(\boldsymbol{A_t}) = \sum_{i=1}^{152} \boldsymbol{\Pi_t} \boldsymbol{\Lambda^{-1}},$$

with $\boldsymbol{\Lambda^{-1}}$ being the inverse of the loadings matrix resulting from the singular value decomposition. In the next step, we derive the trajectories of the WASCSMRs by taking the inverse logistic transform of $logit(\boldsymbol{A_t})$, namely $logit^{-1}[logit(\boldsymbol{A_t})] = \boldsymbol{A_t}$.

---

[11] E.g., PC2 covers cohort shifts within the age groups. For long-term forecasts, the PC should therefore be modeled more detailed. Within the scope of our paper, which is the investigation of short-term fluctuations in mortality, the random walk assumption is sufficient.

The distribution of the difference between the observed WASCSMR and the respective forecasts can then give a probabilistic statement about the actual degree of excess mortality we have observed during a certain period.

The last part of the analysis compares our weekly estimates of excess mortality with the official reported COVID-19 attributed deaths to assess differences between the two data sources. For this, we compute excess mortality numbers between calendar weeks 1 and 25, 2000 – i.e. during the outbreak of COVID-19.

## 4. Results

Figure 5 illustrates the course of the Lee-Carter Index since the beginning of 2019 and its forecast for the first 13 weeks of 2020 with 95% PIs as described in Section 3. Moreover, the dotted red line gives the hypothetical course under the loadings derived from the 2000-2019 data.

*Figure 5. Forecast of Lee-Carter Index for Weeks 1-13, 2020 with 95% PIs and Actual Course*

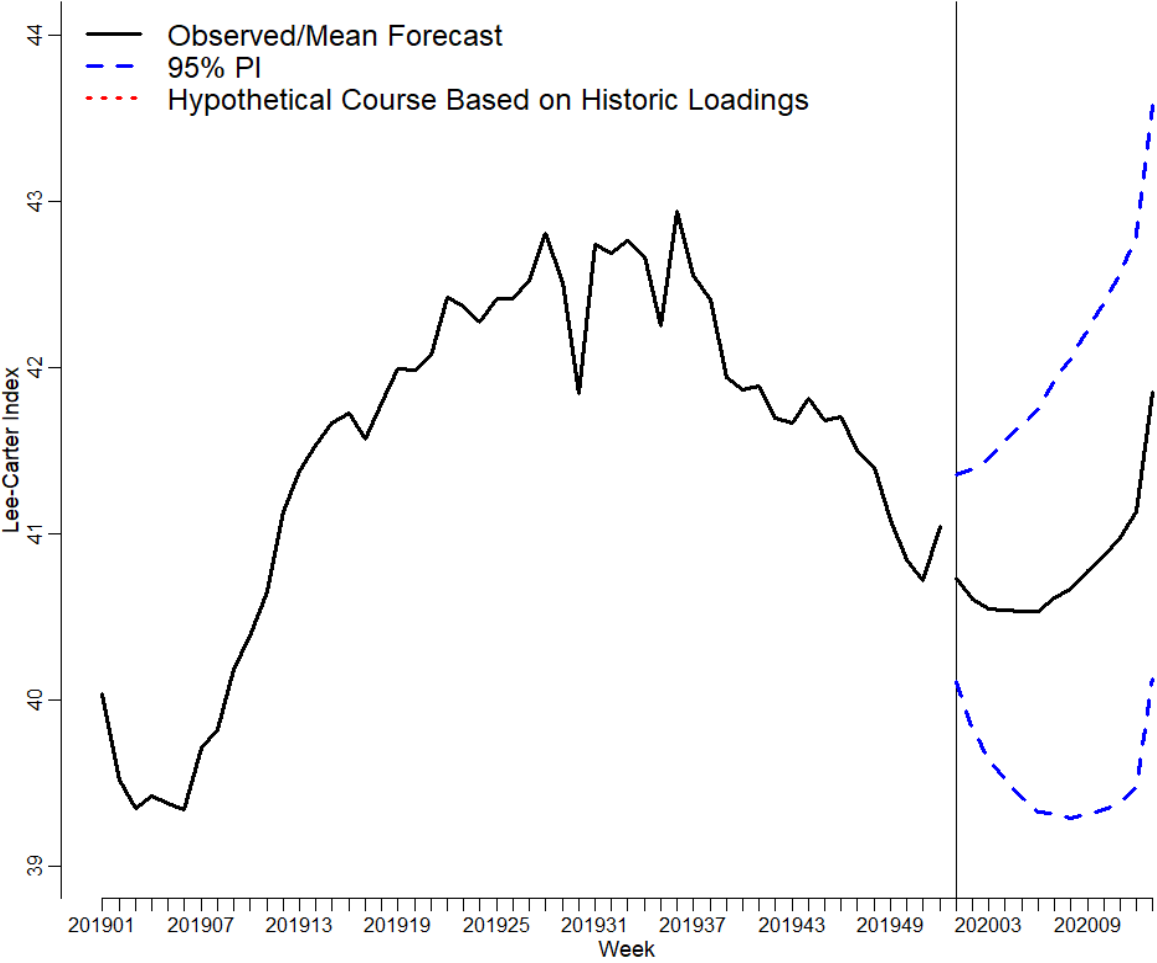

The mortality development oscillates around its mean forecast up to week 10, i.e. the first week of March 2020. After that, it leaves that course and sharply decreases. In week 13, it even falls below the lower bound of the 95% PI. Thus, the international mortality level at this point is statistically significantly higher than the realistic trends which are derived from the previous 20 years of data.

Retransforming the PC forecast to forecasts of the WASCSMRs and multiplying those with the population estimates from the HMD, we derive weekly estimates of death numbers for all sub-groups, which allow for comparison of the observed with the expected mortality level in absolute numbers.

*Figure 6. Observed and Predicted Weekly Death Numbers in 2020 for 18 Study Countries*

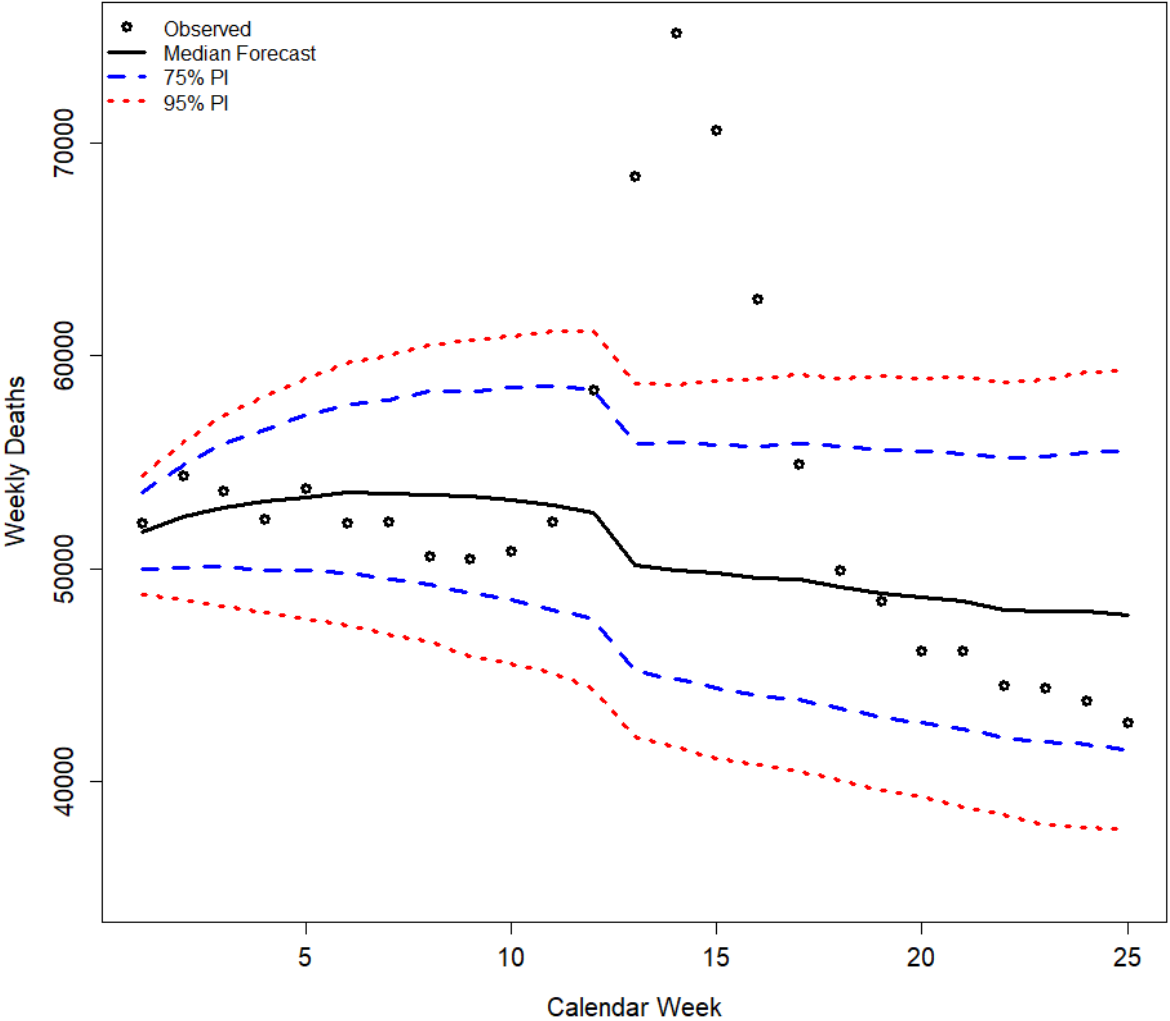

Sources: Human Mortality Database, 2020; Own computation and design

Figure 6 illustrates the overall observed death numbers for weeks 1-25, 2020 for 18 of the study countries[12] compared to the respective predictions with 75% and 95% PIs.

Most of the observations are within the limits of the 75% PI, in week 12 the upper limit is exceeded, while the dot remains within the 95% PI. Between weeks 13 to 16, however, the death numbers exceed the upper limit of the 95% PI.

---

[12] Without Slovenia, as there is no data on deaths for that country after week 13.

Figures 7 and 8 show the analysis stratified by sex and age group.

*Figure 7. Observed and Predicted Weekly Male Death Numbers in 2020 for 18 Study Countries and by Age Group*

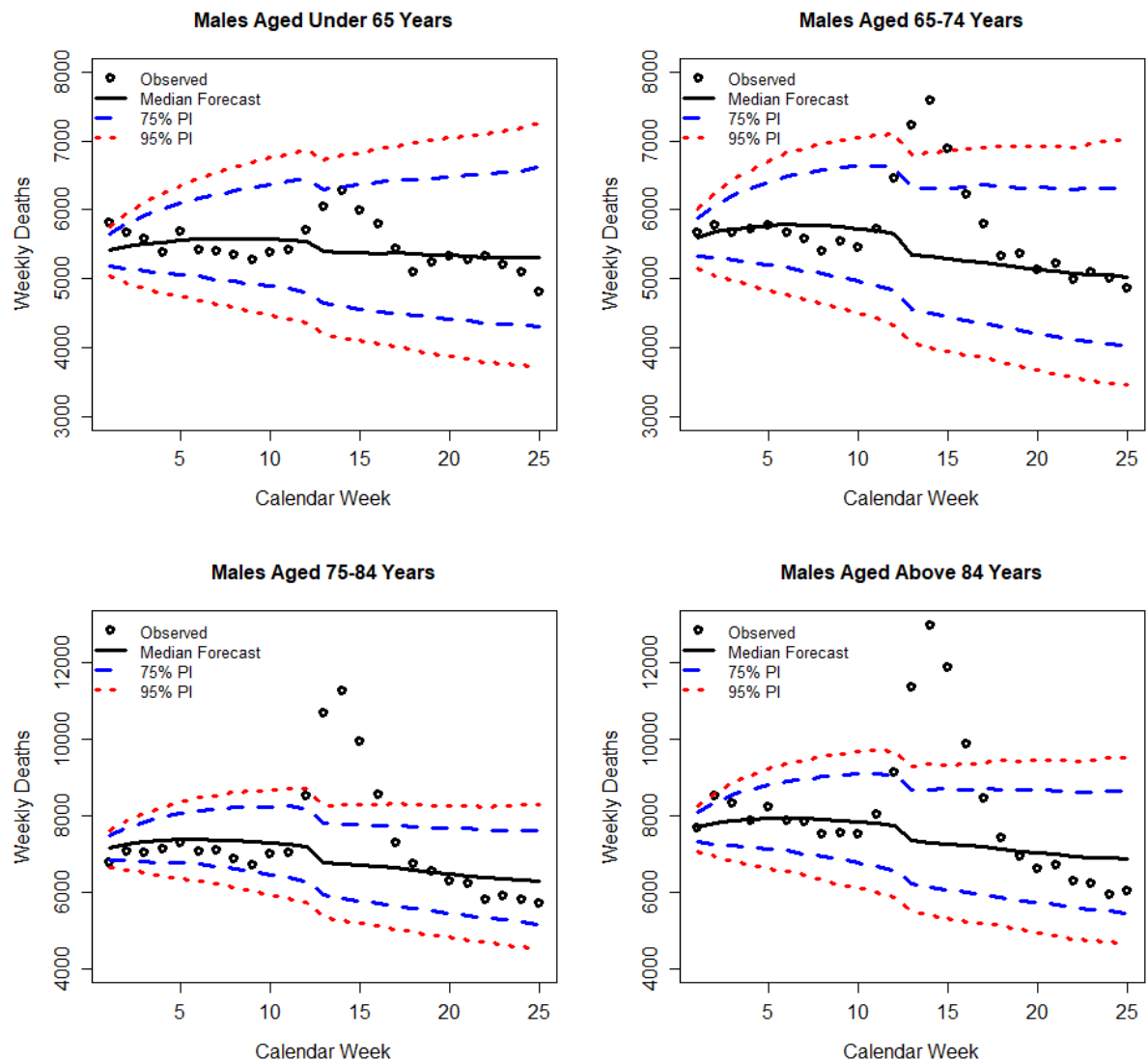

Sources: Human Mortality Database, 2020; Own computation and design

*Figure 8. Observed and Predicted Weekly Female Death Numbers in 2020 for 18 Study Countries and by Age Group*

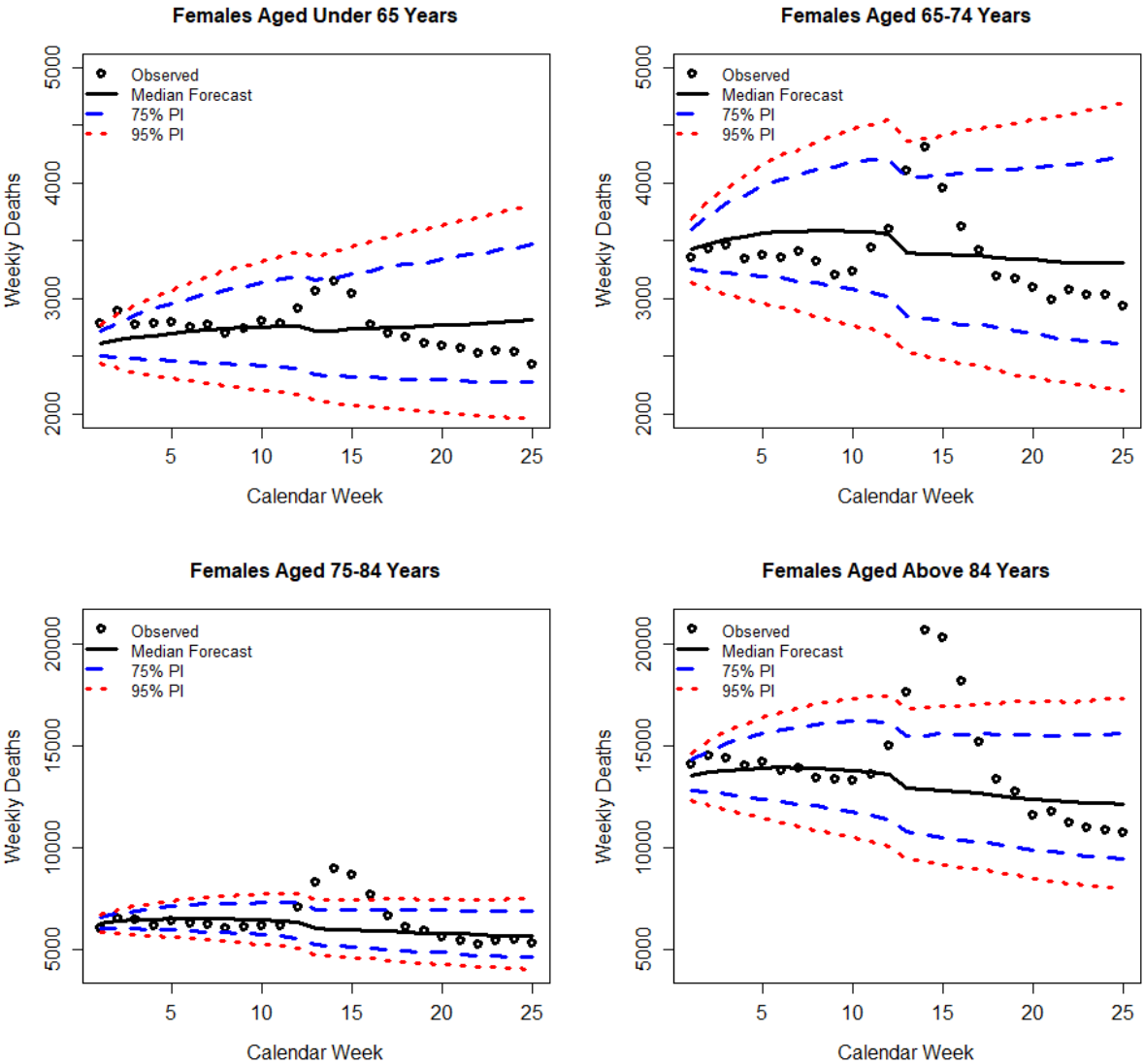

Sources: Human Mortality Database, 2020; Own computation and design

Both sexes and all age groups show a peak in the death numbers between weeks 12 and 17 of 2020. However, a more detailed analysis shows increases beyond the upper limits of the 95% PIs only for the very old age groups. For persons aged 75 and above, we observe a significant increase in mortality for weeks 13-16. For the age group 65-74, the increase is statistically significant for males only. Mortality increases for persons below age 65 since the COVID-19 crisis are not statistically significant for both sexes.

*Figure 9. Observed and Predicted Weekly Death Numbers in 2020 by Country for Countries with Statistically Significant Excess Mortality*

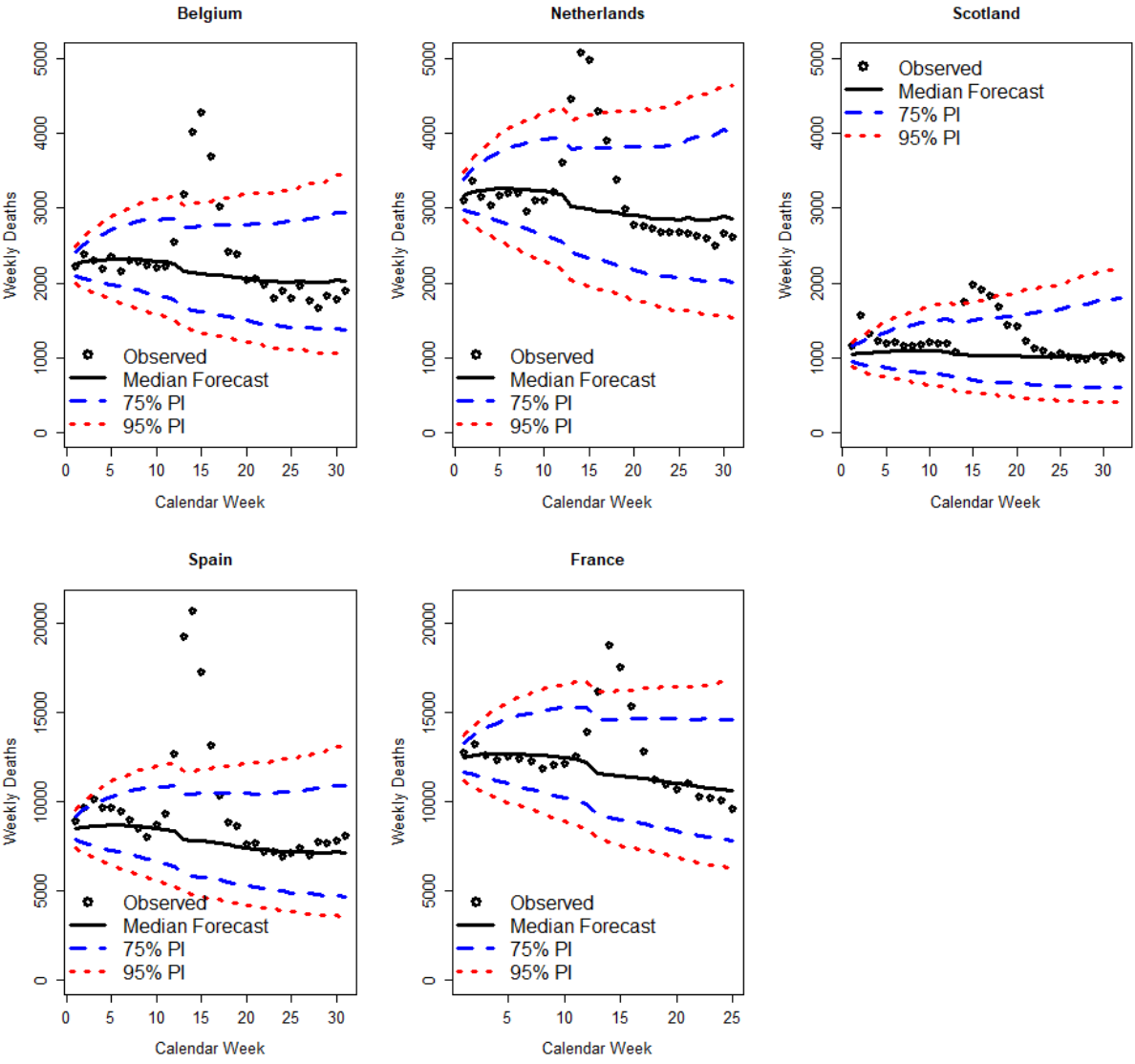

Sources: Human Mortality Database, 2020; Own computation and design

Figures 9-11 investigate the country effect of excess mortality during the COVID-19 crisis. The countries illustrated in Figure 9 show excess mortality between around calendar weeks 13 and 17.

Figure 10 shows the results for the Northern European countries, i.e. the Scandinavian and Balkan countries, without significant excess mortality.

*Figure 10. Observed and Predicted Weekly Death Numbers in 2020 by Country for Northern European Countries without Statistically Significant Excess Mortality*

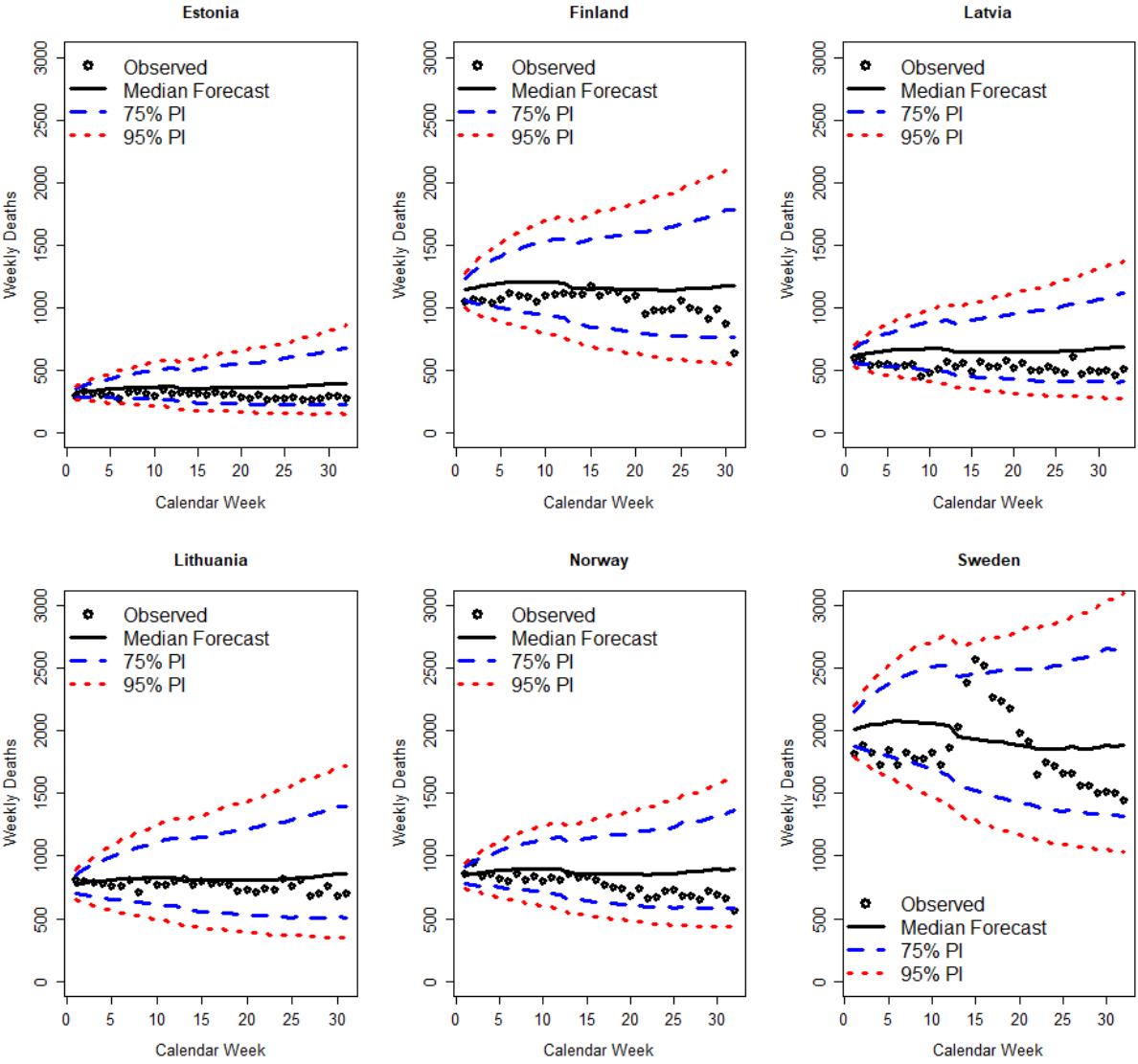

Sources: Human Mortality Database, 2020; Own computation and design

Figure 11 illustrates the analysis for the remaining study countries without significant excess mortality.

*Figure 11. Observed and Predicted Weekly Death Numbers in 2020 by Country for Remaining Countries without Statistically Significant Excess Mortality*

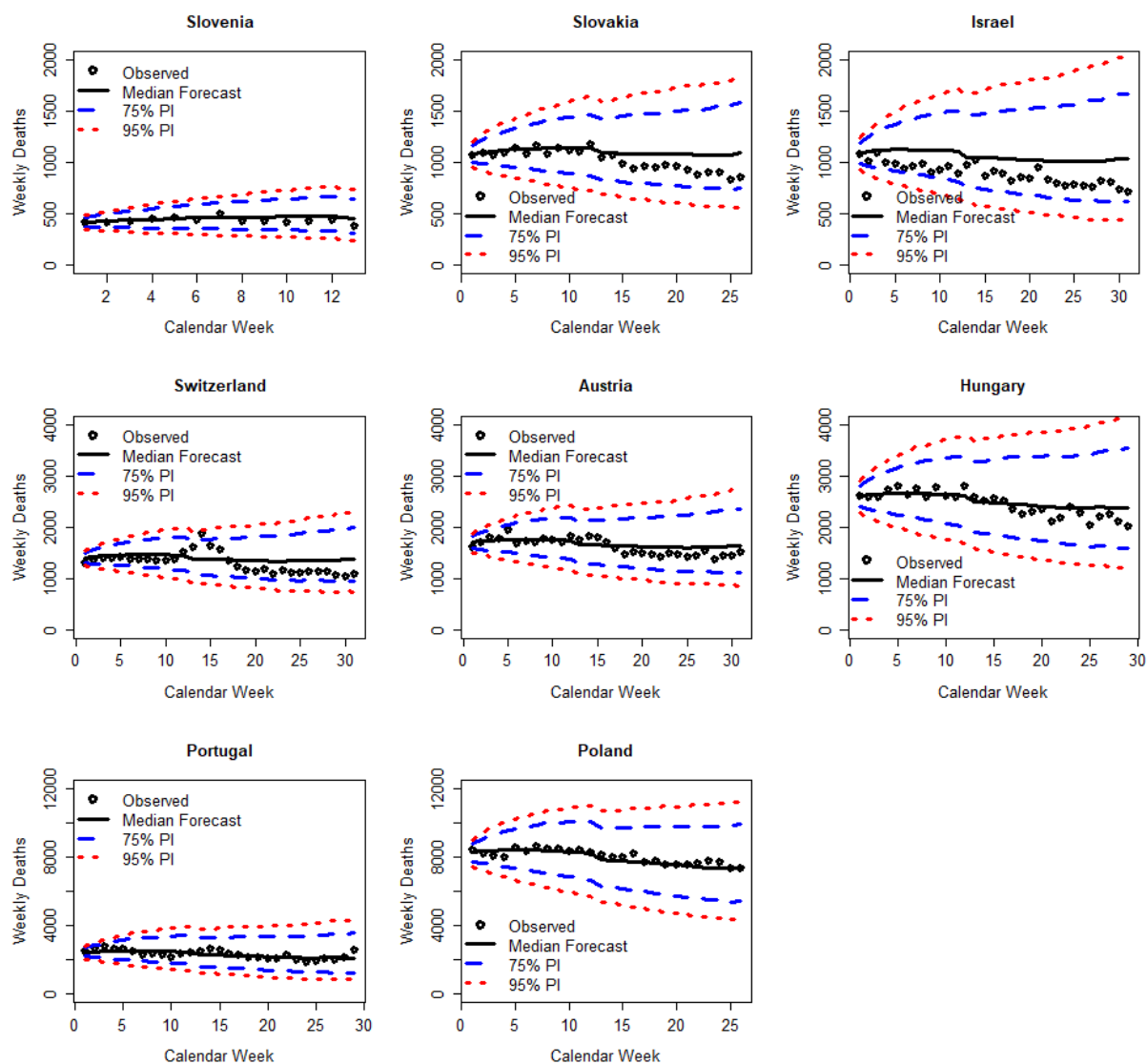

Sources: Human Mortality Database, 2020; Own computation and design

Finally, we investigate how our results relate to official data on COVID-19 associated deaths. Figure 12 illustrates the weekly excess mortality numbers for the 18 study countries from Figure 6, derived from our simulations with 75% and 95% PIs, alongside the official COVID-19 associated deaths, as provided by the ECDC. The bottom panel shows the difference between the excess mortality estimates and the COVID-19 deaths.

*Figure 12. Excess Mortality Distribution with Official COVID-19 Associated Deaths by Calendar Week for 18 Study Countries*

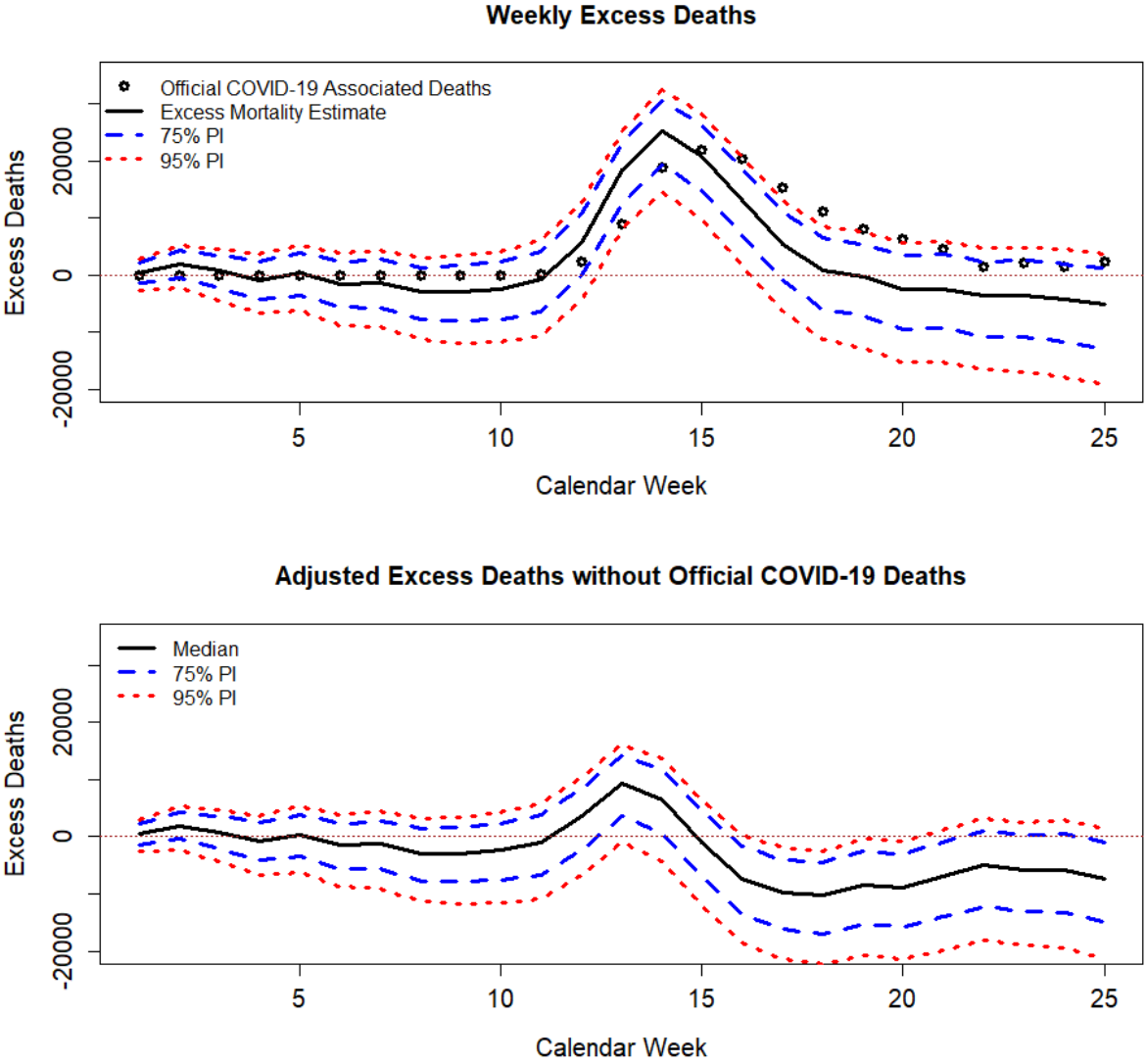

Sources: European Centre for Disease Prevention and Control, 2020; Own computation and design

After subtracting the COVID-19 deaths, there is no significant deviation from the expected death numbers until calendar week 12. Excess mortality in calendar weeks 13 and 14 shows slight tendencies to be augmented even after adjustment to the COVID-19 numbers. After induction of COVID-19 countermeasures, COVID-19 adjusted mortality temporarily was statistically significantly below expectation.

# 5. Discussion

The COVID-19 pandemic has influenced mortality patterns and trends across the world during the first half of 2020. Similar to other analyses (Magnani et al., 2020; Michelozzi et al., 2020; Statens Serum Institut, 2020a), we confirm clear excess mortality in several of those countries with strong infection dynamics during the first half of 2020. Our estimate is however more precise and shows the uncertainty around these estimates based both on the demography of countries as well as long-term mortality trends. Previous approaches do not include the stochasticity in their prediction sufficiently, as they neither consider autocorrelations of the mortality time series (be it death numbers or death rates) nor the cross-correlations among the mortality series in their models, as Appendix C shows. Moreover, some models do not consider the long-term trends in mortality at all, as they simply take the average values of the previous years. Others include the trending behavior, but only for the last four or five years, which does not cover the long-term trending of mortality sufficiently, as we observe decreasing mortality trends in developed countries since at least the early 1970s (Vanella, 2017). The previous approaches to excess mortality estimation therefore systematically underestimate future risk. Forecasts are less certain with increasing distance between the time the forecast was conducted and the time for which the forecast is conducted. This phenomenon is represented by increasing widths of the PIs (e.g., Box et al., 2016; Vanella & Deschermeier, 2020). The literature on excess mortality instead shows constant widths of the intervals. Moreover, not all approaches appear to perform well in the winter season. Statens Serum Institut, 2020a, for instance, shows significant excess mortality in all winter seasons. As excess mortality is the difference between observed and expected deaths, the forecast seems to be systematically misspecified for winter. Our model tries to account for these limitations of previous approaches and can, due to its international perspective, be well implemented for a multi-population analysis of excess mortality.

As the magnitude of our results does not permit us to report everything which could be derived from our model, we restrict the results to one dimension at a time (either demographics or geography by week). Indeed, we derive simulation results for all 152 variables. To illustrate the depth of our analysis, we add the detailed results for all age groups in Spain as an example in Appendix B, since Spain is one of the larger countries in Europe and has witnessed significant mortality due to COVID-19. Moreover, the Spanish COVID-19 data and surveillance are of a relatively high quality.

Our results show that there appears to be a general excess mortality caused by the COVID-19 pandemic, which however affects different age groups and countries heterogeneously. As our study was limited to the countries with sufficiently long time series data, other countries affected strongly by the pandemic, such as Italy, are missing, which limits our conclusions to the countries analyzed here. The excess mortality quantified here is not representative globally. The regional variation is at least partly explained by differing courses of the epidemic as well as different non-pharmaceutical interventions (NPIs) taken nationally or regionally during the study period (Ritchie et al., 2020). Thus it is difficult to quantify the actual attribution of COVID-19 infections on the overall population mortality risk (Chaudry, Dranitsaris, Mubashir, Bartoszko, & Riazi, 2020; Hadjidemetriou, Sasidharan, Kouyialis, & Parlikad, 2020). The mortality due to a specific disease can be addressed by the case fatality risk (CFR) – the risk of death after infection – which, however, is quite vulnerable to bias in outbreaks (Lipsitch et al., 2015). The international CFR estimates for COVID-19 are biased due to demographic characteristics of the cases, time lags between reporting of cases and deaths, and capacities of national healthcare systems, among other unobservable factors. Therefore, assessing international differences in mortality due to this disease without accounting for these characteristics and factors is inadvisable (Backhaus, 2020; Dudel et al., 2020; Vanella et al., 2020).

The classical Lee-Carter model and its extensions, which usually perform exceptionally well in mortality forecasting, might not be applicable in their pure form for the near future, as the long-term overall effect of COVID-19 on the age-specific mortality pattern and its summary measures – such as life expectancy at birth and lifespan inequality – is yet unobserved. Moreover, it is unknown when vaccines will be available to the wide public and how well the vaccines will perform in terms of their effectiveness (Gallagher, 2020). It is highly improbable that the effectiveness of any vaccine will lie at 100% (Folegatti et al., 2020; Lovelace & Higgins-Dunn, 2020), therefore mortality trends derived from the historical mortality data might not be completely representative for the future. As we have not witnessed a similar pandemic in the near past, an adjustment factor to classical Lee-Carter models could be appropriate, transferring them to Bayesian models. The additional information, i.e. the mortality change due to COVID-19, is difficult to assess, however, as the CFRs are biased, as it has been discussed, and the actual prevalence of the disease among the population is unknown. Many cases, who witness only mild symptoms or are completely asymptomatic (Istituto Superiore di Sanità, 2020), will not be detected (Mizumoto, Kagaya, Zarebski, & Chowell, 2020). Moreover, the prevalence estimates are potentially influenced due to a variation of the COVID-19 countermeasures introduced by the different countries and even sub-national geographical units. Taking, e.g., the number of deaths in Spain, illustrated in Figure 9, we observe sharp mortality decreases after the peak in week 13, i.e. the last week of March. In mid-March, Spain introduced national countermeasures to contain the spread of the virus (Hogan Lovells Solutions, 2020), which presumably lead to the mortality decrease after calendar week 13, taking the time lag between an infection and the death of said person of up to two weeks into account (Vanella et al., 2020). As we lack an experimental environment for individual measures, which would be needed for estimating their effect on the virus spread and the mortality, the magnitude of deaths prevented by the countermeasures cannot be quantified. Our stochastic investigation illustrates the poten-tial influence of pure stochasticity on the observed death numbers, which shows that a deterministic inspection of reported death numbers does not give a reliable estimate of the impact of COVID-19 countermeasures, but simply a qualitative orientation. Therefore, the available data does not allow estimation of the mortality level under "normal circumstances", i.e. if we had no active contact reduction measures. Our COVID-19 adjusted estimates have shown, however, that excess mortality in calendar weeks 13 and 14 was exceptionally high, even after taking reported COVID-19 deaths into account. This may be associated either with contemporaneous external factors not associated to the pandemic, to indirect mortality effects of the pandemic, such as surplus mortality through causes due to worse reduced healthcare capacities in overwhelmed healthcare systems (Roberton et al., 2020), or due to bias in the COVID-19 death numbers (Backhaus, 2020) during that time. After the installment of COVID-19 countermeasures in the study countries, our COVID-19 adjusted excess mortality estimates have temporarily been significantly negative. The cause of this is unknown, however direct effects of NPIs on other infectious diseases (including influenza) as well as indirect effects lowering disease burden from other death causes, such as air pollution (Contini & Costabile, 2020) or accidents (Shilling & Waetjen, 2020), is possible. Only cause-specific mortality estimates, including excess mortality, would shine light on these effects – however, the majority of countries included here do not provide death causes in a sufficiently timely manner to allow this. In principle, however, the model presented here would allow stratification by death cause.

An experimental approach might adjust the mortality data to the prevalence rates of active cases among the population. As these are not available for all countries in a similar measure, i.e. in coherent demographic groups (Dudel et al., 2020), well-thought methods for adjusting the available data need to be applied. Adjusting using simulations derived from population-based sero-prevalence studies might be one solution. As this appears to exceed the scope of our investigation, we will not elaborate on that further in this paper. Further studies could consider this, however, for mortality forecasting.

## 6. Conclusions and Outlook

Excess mortality during an epidemic is commonly computed using comparisons of observed death numbers or death rates to predictions of these. Based on an extension of the Lee-Carter mortality model (Lee & Carter, 1992; Vanella, 2017), we introduced a framework for including not only the mentioned autocorrelations of the mortality time series and cross-correlations among the mortality time series into the analysis, but also take long-term trends in the time series into account. We have considered these points in our model by use of a combination of PCA, SARIMA models, and classical time series analysis. Especially the inclusion of international mortality correlations in the model appears to be a crucial aspect in times of pandemics due to the spread of the pathogen over international borders. We have covered the common mortality trends induced by the spread of the virus within our PCA. Moreover, our approach gives an efficient way to conduct multi-population studies on mortality developments. We have illustrated how methods, which are established in demographic forecasting, can enrich the common epidemiological approaches employed in excess mortality studies. Our results identified significant differences in excess mortality among different sub-populations and countries, which could be investigated further.

Our application has illustrated the power of timely and detailed surveillance data on mortality trends in informing health politics in time and provide scientific support for decision-making.

In addition to the case of all-cause mortality covered in this article, our approach could be applied to cause-specific mortality data. This would provide additional insights on mortality patterns related to specific diseases, regardless of the outbreak of an epidemic.

## List of Abbreviations

AIC:                    Akaike's Information Criterion

ARIMA:                  autoregressive integrated moving average

BIC:                    Bayesian Information Criterion

CDC:                    Centers for Disease Control and Prevention

CFR:                    case fatality risk

CI:                     confidence interval

cos:                    cosine

DOHMH:                  New York City Department of Health and Mental Hygiene

ECDC:                   European Centre for Disease Prevention and Control

e.g.:                   exempli gratia

EUROMOMO:               European mortality monitoring

exp(x):                 Euler's number to the power of x

GLM:                    Generalized Linear Model

| HMD: | Human Mortality Database |
|---|---|
| i.e.: | id est |
| $\lambda_i$: | i[th] loading |
| $M_{i,\tau}$: | i[th] logit mortality rate estimate for week $\tau$ |
| MCMC: | Markov Chain Monte Carlo |
| NPI: | non-pharmaceutical intervention |
| PC: | principal component |
| PCA: | principal component analysis |
| PI: | prediction interval |
| $\mathbf{\Pi}_t$: | simulation matrix of the PCs in period $t$ |
| SARIMA: | seasonal autoregressive integrated moving average |
| US: | United States of America |
| $w$: | week |
| WASCSMR: | weekly age-, sex-, and country-specific mortality rate |

# Appendix A. Time Series Numbering

*Table 2. Order of Time Series in the Analysis*

| Number | Country | Sex | Age Group |
|---|---|---|---|
| 1 | Austria | Male | <65 |
| 2 | Austria | Male | 65-74 |
| 3 | Austria | Male | 75-84 |
| 4 | Austria | Male | >84 |
| 5 | Austria | Female | <65 |
| 6 | Austria | Female | 65-74 |
| 7 | Austria | Female | 75-84 |
| 8 | Austria | Female | >84 |
| 9 | Belgium | Male | <65 |
| 10 | Belgium | Male | 65-74 |
| 11 | Belgium | Male | 75-84 |
| 12 | Belgium | Male | >84 |
| 13 | Belgium | Female | <65 |
| 14 | Belgium | Female | 65-74 |
| 15 | Belgium | Female | 75-84 |
| 16 | Belgium | Female | >84 |
| 17 | Switzerland | Male | <65 |
| 18 | Switzerland | Male | 65-74 |
| 19 | Switzerland | Male | 75-84 |
| 20 | Switzerland | Male | >84 |
| 21 | Switzerland | Female | <65 |

| 22 | Switzerland | Female | 65-74 |
| 23 | Switzerland | Female | 75-84 |
| 24 | Switzerland | Female | >84 |
| 25 | Spain | Male | <65 |
| 26 | Spain | Male | 65-74 |
| 27 | Spain | Male | 75-84 |
| 28 | Spain | Male | >84 |
| 29 | Spain | Female | <65 |
| 30 | Spain | Female | 65-74 |
| 31 | Spain | Female | 75-84 |
| 32 | Spain | Female | >84 |
| 33 | Estonia | Male | <65 |
| 34 | Estonia | Male | 65-74 |
| 35 | Estonia | Male | 75-84 |
| 36 | Estonia | Male | >84 |
| 37 | Estonia | Female | <65 |
| 38 | Estonia | Female | 65-74 |
| 39 | Estonia | Female | 75-84 |
| 40 | Estonia | Female | >84 |
| 41 | Finland | Male | <65 |
| 42 | Finland | Male | 65-74 |
| 43 | Finland | Male | 75-84 |
| 44 | Finland | Male | >84 |
| 45 | Finland | Female | <65 |
| 46 | Finland | Female | 65-74 |

| 47 | Finland | Female | 75-84 |
|----|---------|--------|-------|
| 48 | Finland | Female | >84 |
| 49 | France | Male | <65 |
| 50 | France | Male | 65-74 |
| 51 | France | Male | 75-84 |
| 52 | France | Male | >84 |
| 53 | France | Female | <65 |
| 54 | France | Female | 65-74 |
| 55 | France | Female | 75-84 |
| 56 | France | Female | >84 |
| 57 | Scotland | Male | <65 |
| 58 | Scotland | Male | 65-74 |
| 59 | Scotland | Male | 75-84 |
| 60 | Scotland | Male | >84 |
| 61 | Scotland | Female | <65 |
| 62 | Scotland | Female | 65-74 |
| 63 | Scotland | Female | 75-84 |
| 64 | Scotland | Female | >84 |
| 65 | Hungary | Male | <65 |
| 66 | Hungary | Male | 65-74 |
| 67 | Hungary | Male | 75-84 |
| 68 | Hungary | Male | >84 |
| 69 | Hungary | Female | <65 |
| 70 | Hungary | Female | 65-74 |
| 71 | Hungary | Female | 75-84 |

| 72 | Hungary | Female | >84 |
| 73 | Israel | Male | <65 |
| 74 | Israel | Male | 65-74 |
| 75 | Israel | Male | 75-84 |
| 76 | Israel | Male | >84 |
| 77 | Israel | Female | <65 |
| 78 | Israel | Female | 65-74 |
| 79 | Israel | Female | 75-84 |
| 80 | Israel | Female | >84 |
| 81 | Lithuania | Male | <65 |
| 82 | Lithuania | Male | 65-74 |
| 83 | Lithuania | Male | 75-84 |
| 84 | Lithuania | Male | >84 |
| 85 | Lithuania | Female | <65 |
| 86 | Lithuania | Female | 65-74 |
| 87 | Lithuania | Female | 75-84 |
| 88 | Lithuania | Female | >84 |
| 89 | Latvia | Male | <65 |
| 90 | Latvia | Male | 65-74 |
| 91 | Latvia | Male | 75-84 |
| 92 | Latvia | Male | >84 |
| 93 | Latvia | Female | <65 |
| 94 | Latvia | Female | 65-74 |
| 95 | Latvia | Female | 75-84 |
| 96 | Latvia | Female | >84 |

| 97 | Netherlands | Male | <65 |
| 98 | Netherlands | Male | 65-74 |
| 99 | Netherlands | Male | 75-84 |
| 100 | Netherlands | Male | >84 |
| 101 | Netherlands | Female | <65 |
| 102 | Netherlands | Female | 65-74 |
| 103 | Netherlands | Female | 75-84 |
| 104 | Netherlands | Female | >84 |
| 105 | Norway | Male | <65 |
| 106 | Norway | Male | 65-74 |
| 107 | Norway | Male | 75-84 |
| 108 | Norway | Male | >84 |
| 109 | Norway | Female | <65 |
| 110 | Norway | Female | 65-74 |
| 111 | Norway | Female | 75-84 |
| 112 | Norway | Female | >84 |
| 113 | Poland | Male | <65 |
| 114 | Poland | Male | 65-74 |
| 115 | Poland | Male | 75-84 |
| 116 | Poland | Male | >84 |
| 117 | Poland | Female | <65 |
| 118 | Poland | Female | 65-74 |
| 119 | Poland | Female | 75-84 |
| 120 | Poland | Female | >84 |
| 121 | Portugal | Male | <65 |
| 122 | Portugal | Male | 65-74 |
| 123 | Portugal | Male | 75-84 |
| 124 | Portugal | Male | >84 |
| 125 | Portugal | Female | <65 |
| 126 | Portugal | Female | 65-74 |
| 127 | Portugal | Female | 75-84 |
| 128 | Portugal | Female | >84 |
| 129 | Slovakia | Male | <65 |
| 130 | Slovakia | Male | 65-74 |
| 131 | Slovakia | Male | 75-84 |
| 132 | Slovakia | Male | >84 |
| 133 | Slovakia | Female | <65 |
| 134 | Slovakia | Female | 65-74 |
| 135 | Slovakia | Female | 75-84 |
| 136 | Slovakia | Female | >84 |
| 137 | Slovenia | Male | <65 |
| 138 | Slovenia | Male | 65-74 |
| 139 | Slovenia | Male | 75-84 |
| 140 | Slovenia | Male | >84 |
| 141 | Slovenia | Female | <65 |
| 142 | Slovenia | Female | 65-74 |
| 143 | Slovenia | Female | 75-84 |
| 144 | Slovenia | Female | >84 |
| 145 | Sweden | Male | <65 |
| 146 | Sweden | Male | 65-74 |

| 147 | Sweden | Male | 75-84 |
| 148 | Sweden | Male | >84 |
| 149 | Sweden | Female | <65 |
| 150 | Sweden | Female | 65-74 |
| 151 | Sweden | Female | 75-84 |
| 152 | Sweden | Female | >84 |

# Appendix B. Detailed Model Results for Spain

Here, we show the complete results for one of the countries analyzed in our study. Figures 13 and 14 show the observed and predicted number of deaths by age groups and sex in Spain for the weeks 1-30 in 2020. A statistically significant excess mortality is observable in all panels.

*Figure 13. Observed and Predicted Weekly Death Numbers in 2020 in Spain by Sex and Age below 75 Years of Age*

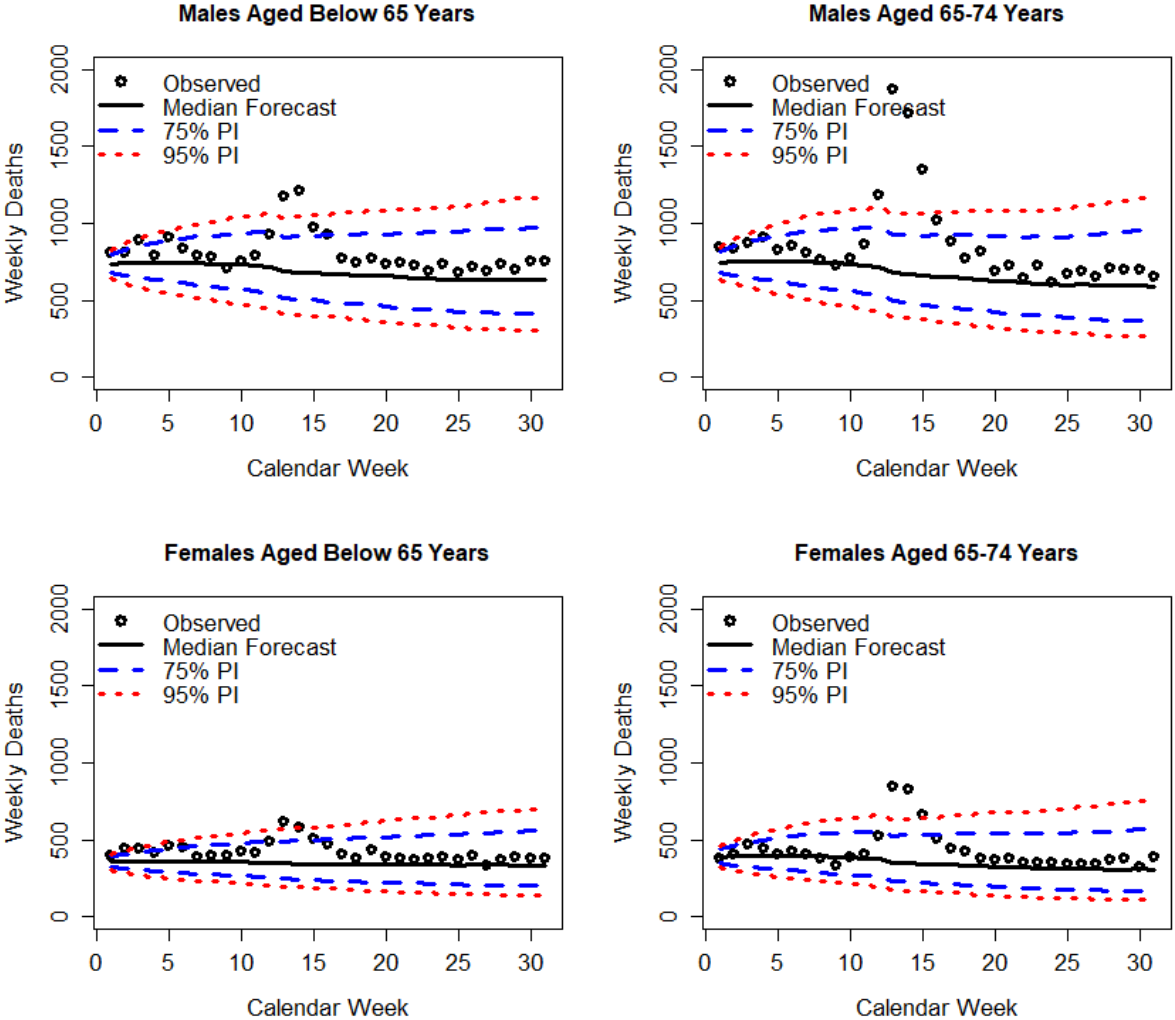

Sources: Human Mortality Database, 2020; Own computation and design

*Figure 14. Observed and Predicted Weekly Death Numbers in 2020 in Spain by Sex and Age above 74 Years of Age*

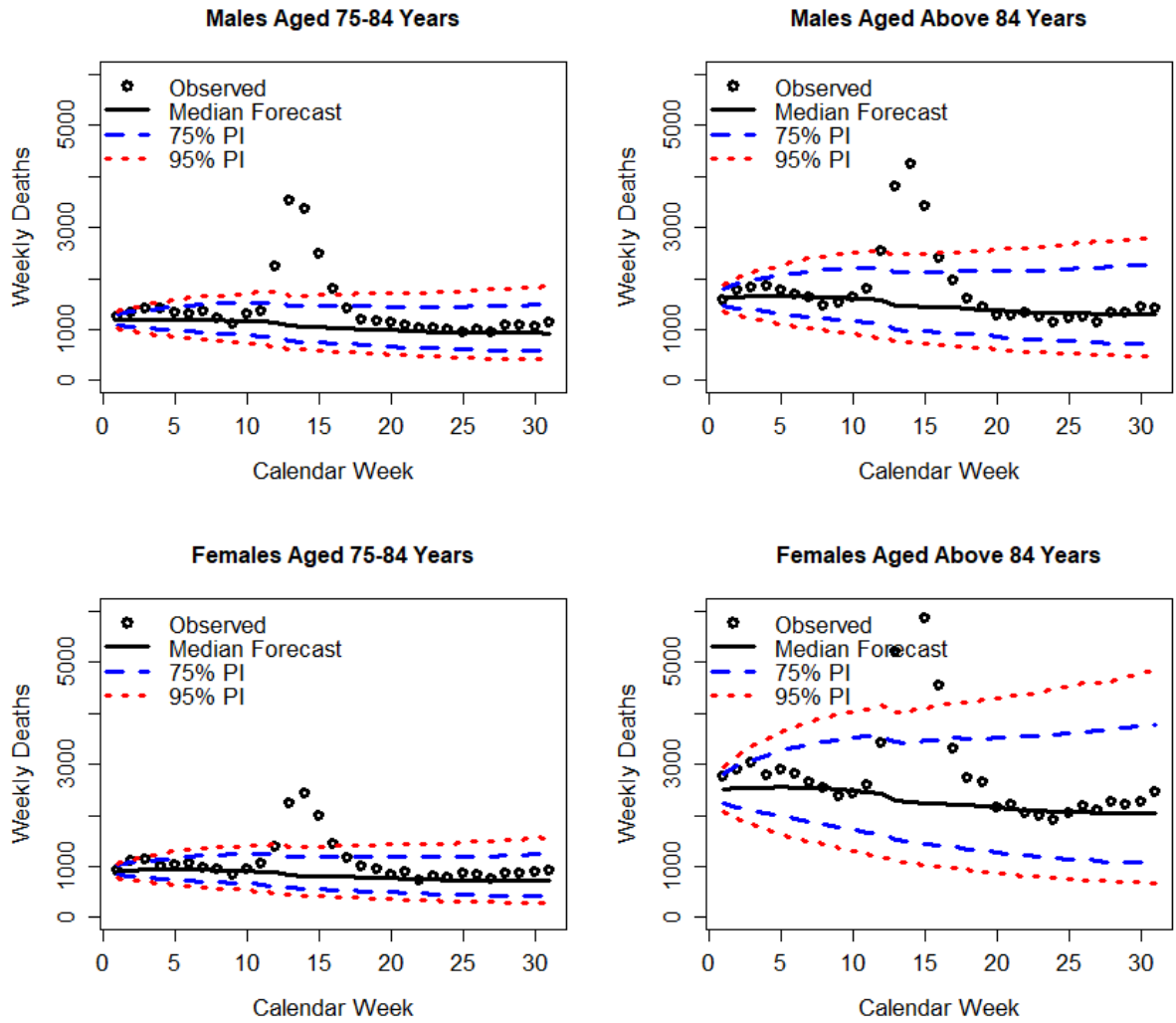

Sources: Human Mortality Database, 2020; Own computation and design

# Appendix C. Selected Approaches for Estimating COVID-19 Excess Mortality

*Table 3. Details on Presented Approaches for Excess Mortality Estimation During COVID-19 Pandemic*

| Study | Geography | Demographics | Data | Method | Main Results | Consideration of Long-term Trends in Mortality | Quantification of Stochasticity | Dealing with Correlations among time series |
|---|---|---|---|---|---|---|---|---|
| Magnani et al. | 4,433 Italian municipalities | No sex distinction<br><br>Two age groups | Daily death numbers from January 1st to 2020 | Mean mortality rates by calendar days for 2015-2020 are computed | Statistically significant increase in mortality rates between early-March and | None | Poisson assumption for observed deaths in 2020 | None |

| | | | April 15th, 2015-2020 Yearly population estimates for January 1st, 2015-2019 Daily Death numbers attributed to COVID-19 by lab testing in 2020 | Daily mortality rate estimates in 2020 are computed and assumed to follow a Poisson distribution Relative risk estimates with 95% CIs are computed relative to the baseline mean rates for 2015-2019 | mid-April, 2020 in Italy Only significant for North and parts of Central Italy Only significant for persons above age 59, except for Lombardia (both age groups significant) | | | |
|---|---|---|---|---|---|---|---|---|

| Michelozzi et al. | 19 Italian cities | Two sexes<br><br>Four age groups | Daily death numbers from January 1st, 2015 to April 18th, 2020 | Mean death numbers by calendar days for 2015-2020 are computed<br><br>Daily death numbers implicitly assumed to be Gaussian, 95% PIs of baseline data are computed<br><br>Comparison of observed death numbers with PIs | Statistically significant excess mortality in Italy between mid-March and mid-April, 2020<br><br>In the North, statistically significant excess mortality for males for all investigated age groups (age 15 and older), for females for age | None | Prediction assumed Gaussian with a standard deviation of past five years | None |

| | | | | | groups 65 and older<br><br>In the Center and South only statistically significant excess mortality for very old females (above age 84) and males (above age 74) | | | |
|---|---|---|---|---|---|---|---|---|
| NYC DOHMH | New York City | None | Daily Death numbers with a lab-confirmed | Daily death numbers are predicted by OLS fit of the Serfling model to | Excess mortality after March 10$^{th}$, with a large | Serfling model | None | None |

| | | | COVID-19 infection in 2020 | daily death numbers of baseline period

Comparison of observed with expected daily death numbers

Comparison of COVID-19 associated deaths with overall excess mortality estimate | share of confirmed or suspected COVID-19 cases | | | |
|---|---|---|---|---|---|---|---|---|
| EURO-MOMO | 22 EU-28 countries | No sex distinction | Weekly all-cause death numbers | Poisson model fit to baseline data | Statistically significant excess mortality in all | Poisson model | PIs from the GLM model | None |
| | 2 German federal states | Seven age groups | since week 1, 2016 | 95% PIs derived from the GLM model<br><br>Comparison of observed weekly death numbers to 95% PIs | countries for age groups 45+ between circa calendar weeks 11 and 19, 2020; death numbers above threshold in calendar weeks 13-15 for persons aged 15-44 years; no significant excess mortality among children | | | |
| | | | | | Significant excess mortality observed in Belgium, France, Ireland, Italy, Netherlands, Portugal, Spain, Sweden, Switzerland, and the UK<br><br>No significant excess mortality in Austria, Denmark, Estonia, Finland, Greece, | | | |
| | | | | | Hungary, Luxembourg, Malta, Norway, and the two German federal states | | | |
| Ours | 19 countries in Europe and the Middle East | Two sexes<br><br>Four age groups | weekly all-cause mortality rate estimates since week 2, 2000 by HMD<br><br>daily official COVID-19 associated | Principal component analysis on all 152 time series of logit mortality rates simultaneously<br><br>OLS fit of logistic SARIMA model to past course of first PC | No significant excess mortality for persons below age 65, slightly significant excess mortality among females aged 65-74 years, statis- | Logistic trend fitted to first principal component time series | SARIMA models and Monte Carlo simulation of principal component time series with nuisance derived from 20 years of past data, leading to stochastic estimates of all mortality rate series | Correlations among age groups and countries completely |

| | | | deaths in 2020 by ECDC | Monte Carlo simulation of weekly forecasts of all PCs

Retransformation of simulations to simulations of age-sex- and country-specific mortality rates

Multiplication of simulations with population estimates for 2020 | tically significant excess mortality among males above age 64 and females above age 74 years in calendar weeks 13-16, 2020

Statistically significant overall excess mortality only observable for Belgium, | | | covered by principal component analysis |

| | | | | | | | | |
|---|---|---|---|---|---|---|---|---|
| | | | | Derivation of non-parametric PIs for forecasts of death numbers<br><br>Comparison of observed all-cause and COVID-19 associated weekly death numbers with PIs of forecasts | France, Netherlands, Scotland, and Spain; results for Sweden and Switzerland inconclusive; no significant excess mortality in Austria, Estonia, Finland, Hungary, Israel, Latvia, Lithuania, Norway, Poland, Portugal, | | | |

| | | | | | Slovakia, and Slovenia | | | |
|---|---|---|---|---|---|---|---|---|